\documentclass[a4paper,10pt,onecolumn]{article}
\usepackage{amsmath,graphicx}
\usepackage[top=1.6cm,left=2.5cm]{geometry}
\usepackage{braket}
\usepackage{datetime}
\usepackage[usenames,dvipsnames]{color}
\definecolor{ReallyDarkBlue}{rgb}{0,0,0.6}

\newcommand{\tobeignored}[1]{}

\newcommand{\eqnref}[1]{Eqn.~(\ref{#1})}
\newcommand{\partialderiv}[2]{\ensuremath{\frac{\partial {#1}}{\partial {#2}}}}

\begin{document}
\title{Wave scattering from rough surfaces for good mirrors}
\author{
	Robert A. Nyman and Benjamin T. Walker\\
	Physics Department, Blackett Laboratory, Imperial College London, SW7 2AZ\\
	email: \texttt{r.nyman@imperial.ac.uk}
	}
\maketitle

This document takes existing derivations of scattering loss from rough surfaces, and makes them more accessible as a tool to derive the total scattering loss from a rough mirror given its true surface profile. It does not contain any new results and is therefore not intended for submission to a scientific journal in the near future.

A rough mirror will diffusively reflect part of an incident wave, limiting the effective specular reflectivity of the mirror. This in turn will limit the finesse of an optical resonator using this mirror. The question we ask is: how does the reflectivity depend on the roughness, in the limit of small roughness?

The reduction of specular reflection is sometimes called the Debye-Waller factor (for people working on electron of X-ray diffraction)~\cite{debye1913interferenz, waller1923frage, lipkin2004physics}. This factor is easy to find (it's even on Wikipedia), but rarely derived. Furthermore, only the root mean squared (rms) value of surface roughness is ever used, which does not tell the whole story, as surfaces may have non-Gaussian deviations with correlations on important length scales. For the record, the reflection from the surface is $e^{-2\sigma^2 k^2}$ where $\sigma$ is the rms roughness, and $k=2\pi/\lambda$ is the incident wavenumber. In the limit of small roughness, the loss from a mirror is $2\kappa^2\sigma^2$. We will see later under what conditions this formula is valid.

The derivation we will use is based off a detailed and well-written book by JA Ogilvy~\cite{Ogilvy} which is almost always out of the library on loan, is out of print, and we can't find any second-hand copies on the internet. Note that nowhere does Ogilvy use the phrase ``Debye-Waller factor''. We outline how this derivation of scattering loss can be used in practice to calculate the scattering loss given a high-precision experimental measure of mirror profile.

\section{First-order perturbation theory}

First, let's write the incident light field, which we'll take to be a scalar plane wave as $\psi^{\rm inc}({\bf r})$. The total field is the sum of incident and scattered fields: $\psi({\bf r}) = \psi^{\rm inc}+ \psi^{\rm sc}$. The surface height is $h(x,y)$. We will make a perturbative expansion around a planar mean surface, which we will set as $z=0$. This expansion will be accurate in first order only if two conditions are satisfied by the height: $k|h(x,y)|\ll 1$ and $|\nabla h(x,y)|\ll1$.

\subsection{Boundary condition}

We will, somewhat arbitrarily (for now), use Dirichlet boundary conditions, so that the total (incident + scattered) wave field is zero at the surface: $\psi({\bf r})|_{z=h}=0$. The equivalent boundary condition on the mean surface $z=0$ is found by Taylor expanding the wave field to first-order:

\begin{align}
	\left(\psi^{\rm inc}({\bf r})\Big|_{z=0} 
	+ h(x,y) \partialderiv{\psi^{\rm inc}({\bf r})}{z}\Big|_{z=0}\right)
	+\left(\psi^{\rm sc}({\bf r})\Big|_{z=0} 
	+ h(x,y) \partialderiv{\psi^{\rm sc}({\bf r})}{z}\Big|_{z=0}\right)
	=0
\end{align}

In perturbation theory we write the scattered field $\psi^{\rm sc} =\psi^{\rm sc}_0 + \psi^{\rm sc}_1 + \cdots $
where the magnitude of $\psi^{\rm sc}_n$ is of order $(kh)^n$. Substituting this in, the boundary condition becomes:
\begin{align}
	\left(\psi^{\rm inc}\Big|_{z=0} 
	+ h(x,y) \partialderiv{\psi^{\rm inc}}{z}\Big|_{z=0}\right)
	+\left(\psi^{\rm sc}_0\Big|_{z=0}+ \psi^{\rm sc}_1\Big|_{z=0}  
	+ h(x,y) \partialderiv{\psi^{\rm sc}_0}{z}\Big|_{z=0}\right)
	=0
\end{align}
The zeroth-order solution drops all terms of order $kh$, so
\begin{align}
	\psi^{\rm inc}\Big|_{z=0} = -\psi^{\rm sc}_0\Big|_{z=0}
\end{align}
Equating the terms which are first-order in $kh$:
\begin{align}
	\psi^{\rm sc}_1\Big|_{z=0} =
	- h\left(\partialderiv{\psi^{\rm inc}}{z}\Big|_{z=0} + \partialderiv{\psi^{\rm sc}_0}{z}\Big|_{z=0}\right)
	\label{eqn:1st order BC}
\end{align}

\subsection{Propagation away from the surface}

At this point the derivation becomes a little more involved, making use of Green's functions to solve an integro-differential equation (see \cite{Ogilvy}, eq (3.10)). We will use Green's identity for two functions $\psi$ and $G$, in a volume $V$ bounded by a closed surface $S$:
\begin{align}
	\int_V \left( \psi \nabla^2 G - G\nabla^2\psi \right) {\rm d}V
	=
	\int_S \left( \psi \nabla G - G\nabla \psi \right) {\bf \cdot }\,{\rm d}{\bf S}
\end{align}
The function $\psi$ is taken to satisfy the wave equation: $\nabla^2\psi + k^2 \psi=0$. The Green's function $G$ satisfies: $(\nabla^2 + k^2)G({\bf r,r_0}) = (\nabla^2 + k^2)G({\bf r_0,r}) = -\delta({\bf r-r_0})$. Here ${\bf r}$ and ${\bf r_0}$ are positions of ``observation'' and ``source'' points respectively. Substituting these properties into the Green's identity:
\begin{align}
	-\int_V \psi({\bf r_0})\delta({\bf r-r_0}){\rm d}V_0
	=
	\int_S \left( \psi \nabla_0 G - G\nabla_0 \psi \right) {\bf \cdot }\,{\rm d}{\bf S_0}
\end{align}
where the subscript $0$ refers to the source co-ordinates. The left-hand side picks out  $\psi({\bf r})$, so we come to an expression for the scattered field in terms of the field at the surface:
\begin{align}
	\psi^{\rm sc}({\bf r})=\int_S \left( \psi \nabla_0 G - G\nabla_0 \psi \right) {\bf \cdot }\,{\rm d}{\bf S_0}
\end{align}
Applying the boundary condition of \eqnref{eqn:1st order BC}, we obtain for the scattered field some distance from the surface:
\begin{align}
	\psi^{\rm sc}_1({\bf r}) = -\int_{S_M} h(x_0,y_0)\left(
		\partialderiv{\psi^{inc}}{z} + \partialderiv{\psi^{\rm sc}_0}{z}
	\right)
	\partialderiv{\widetilde{G}({\bf r,r_0})}{z_0} \,{\rm d}S_M({\bf r_0})
\end{align}
where $S_M$ is a source term on the mean surface $z_0=0$. This assumes that $\partialderiv{\psi^{\rm sc}_1}{z_0}=0$ or $\simeq0$, which is fair because it is of order $h$, so small.

The function $\widetilde{G}$ is in fact the Green's function for the half-space (the incoming-wave side of our mirror) which is given by:
\begin{align}
	\widetilde{G}({\bf r,R}) = \frac{e^{{\rm i}k|{\bf r-R}|}}{4\pi|{\bf r-R}|}
		- \frac{e^{{\rm i}k|{\bf r-R'}|}}{4\pi|{\bf r-R'}|}
\end{align}
The vectors ${\bf R}=[X,Y,Z]$ and ${\bf R'}=[X,Y,-Z]$ are mirror reflected. 

To first order, the scattered field, which is proportional to the height, averages to zero, and so does not change the coherently-scattered specular reflection. However, the first-order, incoherent, diffuse reflection intensity depends on the square of the field: $\braket{I_1} = \braket{{\psi^{\rm sc}_1}^* \,\psi^{\rm sc}_1}$. Using ${\rm d}S_M({\bf r})={\rm d}x\, {\rm d}y$:
\begin{align}
	\braket{I_1({\bf r})} 
	= &{\psi^{\rm sc}_1({\bf r})}^*\psi^{\rm sc}_1({\bf r})\nonumber\\
	= &
	\iint \,{\rm d}x_0\, {\rm d}y_0 {\rm d}x_1\, {\rm d}y_1\,\,
		h(x_0,y_0)h(x_1,y_1)\,\times
	\\
	&\hspace{2ex}
		\left[\partialderiv{{\psi^{inc}({\bf r_1})}^*}{z} + \partialderiv{{\psi^{\rm sc}_0({\bf r_1})}^*}{z}\right]
		\left[\partialderiv{\psi^{inc}({\bf r_0})}{z} + \partialderiv{\psi^{\rm sc}_0({\bf r_0})}{z}\right]
		\partialderiv{{\widetilde{G}({\bf r,r_1})}^*}{z_1}
		\partialderiv{\widetilde{G}({\bf r,r_0})}{z_0} 
	\nonumber
\end{align}
The factor $h(x_0,y_0)h(x_1,y_1)$ is related to the correlation of surface deviations from the mean plane, and is proportional to the square of their magnitude.

\subsection{Scattered intensity for a plane incident wave}

An incident plane wave is of the form $\psi^{\rm inc}({\bf r}) = e^{{\rm i}{\bf k_{inc}\cdot r}}$. We will carefully define a co-ordinate system, as in Ogilvy~\cite{Ogilvy}, p. 42, but we will specialise immediately to the case of normal incidence. If $\theta_1$ is the angle between the incident wave and the normal to the plane, $\theta_1=0$ so ${\bf k_{inc}}= k\,{\bf e_z}$. The scattered wave direction is specified by the two angles $\theta_2$, the angle between the scattered wave direction and the normal to the mean plane, and $\theta_3$, the angle between the scattered wave direction and the $x$-axis. The scattered wavevector is 
\mbox{${\bf k_{sc}} = k({\bf e_x}\sin\theta_2\cos\theta_3 + {\bf e_y}\sin\theta_2\sin\theta_3 + {\bf e_z}\cos\theta_2)$}

In the far field, $kr\gg 1$ (many wavelengths away from the surface) and $r\gg r_0$ (many times further away than the size of the surface).\footnote{From this point on, we have not checked Ogilvy's calculations, as they are rather involved.} This means that the Green's function can be approximated as $\widetilde{(G)} \simeq \frac{-{\rm i}k e^{{\rm i}kr}}{2\pi r} e^{-{\rm i}{\bf k_{sc}\cdot r_0}} \cos \theta_2$. The average scattered intensity is then:
\begin{align}
	\braket{I_1} = \frac{4k^4 \cos^2\theta_2}{r^2}A_M P(kA,kB)
\end{align}
Stationary surface roughness has been assumed (roughness looks the same at different places on the surface). $A_M$ is the surface area of the plane (the mirror), and $P(k_1,k_2)$ is the surface roughness power spectrum as defined by (\cite{Ogilvy}, Eqn. (2.15)):
\begin{align}
	P(k_x,k_y) = \lim_{A_M\rightarrow \infty} \, \frac{1}{4\pi^2 A_M}
		\left|\int h(x,y) \,e^{{\rm i}{(k_xx + k_yy)}}\,\, {\rm d}x\,{\rm d}y\,\right|^2
	\label{eqn: power spectrum}
\end{align}
where ${\bf k}=(k_1,k_2)$.
The scaling factors $A$ and $B$ are $A=-\sin\theta_2\cos\theta_3 \text{ and } B=-\sin\theta_2\sin\theta_3$

One way of interpreting these formulae is that each spatial frequency component of the surface roughness diffracts light into a different direction.

\subsection{Total scattered power}

Starting from the preceding equations, we see that at a distance $r$ the total scattered power is
\begin{align}
	P^{\rm sc} =& \int_{\theta_3=0}^{2\pi}\int_{\theta_2=0}^{\pi/2} \braket{I_1} r^2 \sin\theta_2 {\rm d}\theta_2\,{\rm d}\theta_3\\
	=& \int_{\theta_3=0}^{2\pi}\int_{\theta_2=0}^{\pi/2} {4k^4 \cos^2\theta_2 \sin\theta_2}A_M P(kA,kB)  {\rm d}\theta_2\,{\rm d}\theta_3
\end{align}
The total incident light power is $A_M$ in these units, so the loss coefficient is:
\begin{align}
	\label{eqn:loss}
	\Lambda = \int_{\theta_3=0}^{2\pi}\int_{\theta_2=0}^{\pi/2} {4k^4 \cos^2\theta_2 \sin\theta_2}P(kA,kB)  {\rm d}\theta_2\,{\rm d}\theta_3\\
	\text{ with } A=-\sin\theta_2\cos\theta_3 \text{ and } B=-\sin\theta_2\sin\theta_3\nonumber
\end{align}
and the surface noise power spectrum is defined in \eqnref{eqn: power spectrum}

It is worth noting that the lowest spatial frequencies diffract to very small angles (nearly back into the coherent specularly-reflected beam), but the angular factor $\cos^2\theta_2\sin\theta_2$ tends to zero, so these components do not contribute. Likewise, for $\theta_2\simeq\pi/2$, corresponding to spatial noise frequencies close to the incident wavenumber, there is almost no contribution due to the same angular factor. 

Spatial noise at wavelengths shorter than the incident wavelength do not contribute to loss, since $\theta_2>\pi/2$ would imply forward scattering, not reflection. The problem is periodic, so aliasing of even higher frequency noise may cause loss, for $3\pi/2<\theta_2<5\pi/2$ and further intervals of $2\pi$, although I'm sure some other phenomenon intervenes.

\begin{figure}[hbt]
	\centering
	\includegraphics[width=5cm]{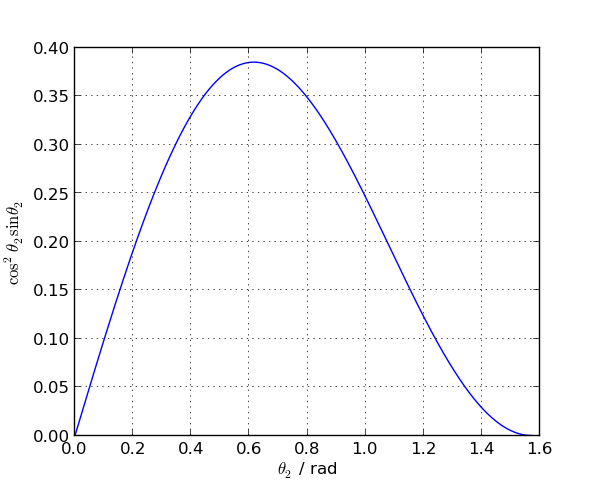}
	\caption{Angular factor in integral relation between loss and power noise spectrum of surface roughness}
	\label{fig:angular factor}
\end{figure}

The peak of the angular factor is found at $\theta_2 \simeq 0.46$~radians. Therefore, the spatial frequencies that contribute most to the loss are $k\sin\theta_2 \simeq 0.45k$, i.e. about twice the incoming wavelength as shown in Fig. \ref{fig:angular factor}.

The exact form of this angular factor depends on the boundary condition applied. We have given the values for scalar-wave Dirichlet ($\psi=0$) boundary conditions; scalar-wave von Neumann and polarised waves give different forms. However, they always include a factor $\sin\theta_2$ for simple geometric reasons, so that long-wavelength roughness doesn't cause much loss. Likewise, since $\sin\theta_2$ can't be bigger than 1, wavenumbers bigger than $k$ don't contribute at all.

\subsubsection{Electromagnetic waves}

Ogilvy treats the case of electromagnetic waves incident on a perfectly conducting surface. The result is a formula much like \eqnref{eqn:loss}, but with a different angular factor. 
\begin{align}
	\Lambda = \int_{\theta_3=0}^{2\pi}\int_{\theta_2=0}^{\pi/2} {k^4 \Phi(\theta_2,\theta_3)\sin\theta_2}P(kA,kB)  {\rm d}\theta_2\,{\rm d}\theta_3
\end{align}
Ogilvy's equation (5.7) shows that $\Phi = 4\cos^2\theta_2$, at least so long as the polarisation is unchanged by the scattering, my previous conclusions remain valid, and \eqnref{eqn:loss} can be directly applied. Dielectric mirrors also correspond to a Dirichlet boundary, since the reflection comes from exactly destructive interference by a stack of partially scattered waves.

\subsection{Special case: Gaussian noise with Gaussian correlation distribution}
Let us suppose that the power noise spectrum is an isotropic Gaussian, peaked at very long wavelengths:
\begin{align}
	P(k_x,k_y) = \sigma^2 \frac{\lambda_C^2}{\pi}  e^{-\lambda_C^2 (k_x^2 + k_y^2)}
\end{align}
where $\lambda_C$ is a correlation length and $\sigma$ is an amplitude parameter, and the normalisation condition is $\iint P(kx,ky) {\rm d}k_x {\rm d}k_y = \sigma^2$. Plugging this into \eqnref{eqn:loss}:
\begin{align}
	\Lambda =& \int_{\theta_3=0}^{2\pi}\int_{\theta_2=0}^{\pi/2} 
		{4k^4 \cos^2\theta_2 \sin\theta_2}
		\sigma^2 \frac{\lambda_C^2}{\pi}  
			e^{-\lambda_C^2 \sin^2\theta_2 (k^2\cos^2\theta_3 + k^2\sin^2\theta_3)}
		{\rm d}\theta_2\,{\rm d}\theta_3\\
	=& 8k^4 \sigma^2 \lambda_C^2
		\int_{\theta_2=0}^{\pi/2} 
			{\cos^2\theta_2 \sin\theta_2}\,e^{-\lambda_C^2 k^2\sin^2\theta_2 }
				{\rm d}\theta_2
\end{align}
We perform the final integral in Mathematica, giving the Dawson $F$ function ($D_+$), which is never bigger in magnitude than 0.5, and for large arguments $\lambda_C^2 k^2 \gg 1$ tends to small values. 
\begin{align}
	\Lambda =& 8k^4 \sigma^2 \lambda_C^2
		\left[ \frac{1}{2 \lambda_C^2 k^2}  - \frac{D_+(\lambda_C k)}{2 \lambda_C^3 k^3}\right]
\end{align}

For $\lambda_C k \ll 1$, $D_+(\lambda_C k) \sim \lambda_C k$ and therefore $\Lambda \to 0$ as expected. For $\lambda_C k \simeq 1$ we get $D_+(\lambda_C k) \simeq 0.5$ and $\lambda_C^2 k^2 \simeq \lambda_C^3 k^3$, therefore recovering the expected Debye-Waller factor $\Lambda \simeq 2k^2 \sigma^2$. For $\lambda_C k \gg 1$, $D_+(\lambda_C k) \to 0.5$ and an extra factor of 2 appears giving $\Lambda \simeq 4k^2 \sigma^2$. Note that in the limit $\lambda_C k \gg 1$, the scattering loss becomes independent of $\lambda_C$. The physical origin of this additional factor of 2 in the limit $\lambda_C k \gg 1$ compared to the usual Debye-Waller factor is not immediately clear to us, but 
we do not discuss this discrepancy in detail here.


Note that this derivation assumes a Gaussian power spectral density for the roughness, but does not strictly assume Gaussian-distributed heights. The surface roughness should be measured with a length scale which is comparable to, but not shorter than, the incident wavelength.

\subsection{Numerical Implementation}

From Equations~\ref{eqn: power spectrum} and~\ref{eqn:loss}, calculating the expected loss is from a high-precision measurement of a mirror profile is straightforward. Equation~\ref{eqn: power spectrum} is simply the Fourier transform of the mirror surface profile for which numerically-efficient packages are available in most programming languages. Finally, we express $\theta_2$ and $\theta_3$ in terms of $k_x$ and $k_y$, noting that $kA=k_x$ and $kB=k_y$. We use the Jacobian to transform the integral over $(\theta_2,\theta_3)$ to an integral over $(k_x, k_y)$, yielding

\begin{align}
	\label{eqn:loss_2}
	\Lambda = \iint{4k \sqrt{k^2-k_x^2-k_y^2}(k_x^2+k_y^2)P(k_x,k_y)}  {\rm d}k_x\,{\rm d}k_y
\end{align}

where $P(k_x,k_y)$ is the two-dimensional Fourier transform of the surface profile. It is again clear that there is very little scattering loss from roughness with small wave-vector (this roughness simply appears as a global shaping of the mirror surface, allowing good reflection), and no scattering for large wavevectors of roughness (the incident light has too long a wavelength to probe roughnesses on short scales).

\section{The assumptions used}
\begin{itemize}
	\item Small surface height roughness compared to the incident wavelength.
	\item Gradients of surface height are small.
	\item Roughness looks statistically the same on all parts of the mirror.
	\item Mirror much larger than the incident wavelength.
	\item Observation in the far field, i.e. at a distance much larger than the incident wavelength, roughness correlation length and mirror size.
	\item Perfect reflection from a perfectly planar surface of the mirror material.
\end{itemize}
We've not thoroughly checked Ogilvy's formula relating to my equations 8, 9, 11 and 16, but we see no reason to doubt Ref. \cite{Ogilvy}. 

\bibliographystyle{unsrt}
\bibliography{scattering_refs.bib}

\end{document}